# A QoS based Routing Approach using Genetic Algorithms for Bandwidth Maximization in Networks

T R Gopalakrishnan Nair, Kavitha Sooda[*] and R Selvarani[**]

**Saudi Aramco Endowed Chair**, Technology and Information Management, PMU. KSA
**Vice President Research**, DSI, Bangalore-78,( Sabbatical), India
Phone: +91-80-42161766
E-mail: trgnair@ieee.org

[*] Nitte Meenakshi Institute of Technology,
 P B No.6429, Gollahalli, Govindapura,
 Yelahanka, Banglore-64, India.
 E-mail: kavithasooda@gmail.com

[**] King Abdul Aziz university Saudi Arabia
 M S Ramaiah Institute of Technology
 Bangalore – 560054 (Sabbatical), India
 Phone: +91-90356 63245
 E-mail: selvss@yahoo.com

[*] Corresponding Author

**Abstract:** This paper addresses the path selection problem from a known source to the destination in dense networks. The proposed solution for route discovery uses the genetic algorithm approach for a QoS based network. The multi point crossover and mutation helps in determining the optimal path and alternate path when required. The input to the genetic algorithm is a learnt module which is a part of the cognitive router that takes care of four QoS parameters. Here the set of nodes selected for routing is determined by delay, jitter and loss. On this graded surface of nodes selected, the bandwidth parameter is considered for path selection. The aim of the approach is to occupy the maximized bandwidth along the forward channels and minimize the route length. The population size is considered as fixed nodes participating in the network scenario, which will be limited to a known size of topology. The simulated results show that by using genetic algorithm (GA) approach the probability of convergence to shortest path is higher.

**Keywords:** routing; cognition; genetic algorithm; multipoint cross over; bandwidth; QoS; mutation; optimal path; intelligent system; elitism;

**Biographical Notes: T.R.Gopalakrishnan Nair** holds M.Tech. (IISc, Bangalore) and Ph.D. degree in Computer Science. He has 3 decades experience in Computer Science and Engineering through research, industry and education. He has published several papers and holds patents in multi domains. He is winner of PARAM Award for technology innovation. Currently he is the Saudi Aramco Endowed Chair, Technology and Information

Management, PMU, KSA and VP of Research and Industry in Dayananda Sagar Institutions, Bangalore, (Sabbatical), India.

**Kavitha Sooda** holds M.Tech in Computer Scinece and Engineering. She has ten years of teaching experience and pursuing her Ph. D from JNTU, Hyderabad. Her interest includes routing techniques, QoS application, cognitive networks and genetic algorithms. Currently she works as Assistant Professor at Nitte Meenakshi Institute of Technology, Bangalore, India.

**R.Selvarani** holds M. Tech., Ph.D. degree in Computer Science and Engineering. She has 20 years of experience in teaching including research. She has published several research papers in computer science and holds two patents. She has been awarded best teacher award twice in various institutions. Currently she is working as Professor and Head of Department of Computer Science and Engineering, MSRIT, Bangalore, (Sabbatical), India.

# 1 Introduction

Computer networks in general have been performing the tasks that were defined by user requirements. With many specialized tasks being introduced, the network has developed a best effort approach in performance by providing QoS. This model has served well till date, but as the end-user applications augments their requirements and a set of similar scenarios repeating in the network, the network tends to increase its complexity and complicate itself in order to deliver the quality in service to the end user. This aspect has generated lot of research interest in active networks. The main objective of this new research field has been to enable the network to evolve to be intelligent. Despite the several new technologies which have been developed, the management and maintenance of the network will still be a low-level task lying at the hands of network administrators. This is due to the fact that the network will not be *aware* of its state and its needs, and also will not have *knowledge* of its goals and the method for 'how' to achieve them. Finally, to be successful in its objectives, it should l be able to *reason out optimal path*. Such properties would render the network to be autonomic and self-manageable by being cognitive. Thus cognitive networks refer to a futuristic research area that envisions combining current active network technology with mature AI methodologies (such as reinforcement learning, expert system, neural networks), with a sole aim to build a context of reasoning and cognition in the network fabric. This is expected to enable the construction of dependable, resilient, manageable, self-configurable and finally survivable networks. Much work has been derived for the learning and planning of the cognitive system. In this cognitive system routing plays a vital role. Routing is a process of forwarding the data from a known source to a desired destination. In this process, the data may travel through several intermediate paths, where there is a need to select the best possible optimal nodes to forward the data. This selection of nodes is done to achieve a high performance of the network. There are many existing work done in this area of route discovery which are discussed in the literature by Gelenbe (2006), Viswanatha (2009) and Nakamura (2000). The existing algorithms have found the optimal paths considering either one or two QoS parameters or hop counts or cost as the deciding factor for route selection. The proposed work assumes four QoS parameters such as bandwidth, delay, jitter and loss which act as the input to the GA approach and bandwidth availability at the links for finding the optimal path.

In order to make the present network systems to be intelligent there is a need for an open platform discussion on cognitive experiments. A common building block has been proposed by Thomas (2008). Bandwidth availability has been determined by multi hop analysis as in Loguinov (2008). The Setting up of the geographical layout for cognitive networks is described in Gao (2009). A model which combines a reconfigurable core and control systems along with genetic algorithms for cognitive functionality has been dealt in Doyle (2007). The security aspects are dealt in Prasad (2008), which discuss the research challenges for security in cognitive networks. Among the major key security aspects which are dealt in Prasad (2008), the communication control channel jamming congestion is automatically avoided by our approach as data is forwarded based on the availability of bandwidth at the given link. The proposed work follows the architecture model as in Noelan (2006) where as the deciding module has been dealt with reasoning capability.

The objective of this paper is to find an efficient solution for end-to-end delivery which involves geographical intelligence and multiple router integration at large distance. However, we handle several layers of routers to prove GA based selection of channels which can be used in cognitive routing. In the recent past, genetic algorithm has found its application in route selection algorithms as discussed in Gelenbe (2006). The fact that GA learns with the system and applies it to the environment makes it better than the existing algorithms. Thus GA has been used for path selection in the ever changing network scenario. This will lead to better network performance as compared to the existing predetermined routing.

Routing is a process of forwarding the data from a known sender to the particular receiver. The nodes are initially segregated based on delay, jitter and loss which act as the input to GA approach and bandwidth availability at the links for finding the optimal path.

The paper is organized as follows: Section 2 addresses the Intelligence in routing. Section 3 gives the background work on QoS and reasoning. The fitness function, crossover and mutation methods are shown in Section 4. The simulation results are shown in Section 5, and the conclusions and future works are dealt with in section 6.

## 2 Intelligence in Routing

Routing algorithms play an important role in the path determination process. A good routing algorithm should be able to find an optimal path and it must be simple, stable, converge rapidly and must remain flexible. There exists a lot of routing algorithm which have been developed for specific kind of network as well as for general routing purpose. The existing algorithms are either, table driven or demand-driven protocols. These algorithms have been used for different applications depending on their specification. Still dynamics of nodes, hidden terminals, power aware routing; location-aid routing remains a challenge other than the QoS and multicasting. With all this limitations, the protocols have been supporting the existing users of the internet. But the current trend in the usage of the internet shows that these protocols will not be self-sustainable for meeting the requirement of fast end-to-end delivery.

This would require a knowledge base reference where the node would learn, remember and recollect whether it has come across such a scenario previously. This requires the network to be composed of elements that, through learning and reasoning, dynamically adapt to varying network conditions in order to optimize end-to-end performance. This will lead to meeting the requirements of the network as a whole, rather than the individual network components as discussed in Nair (2008). This is where cognitive networks play an important role. Here cognition is used to observe the forward channels and making behavioral adjustments to seek the best path. It receives the feedback from other nodes while learning which is based on the QoS parameters. The elements of this cognitive network are capable of assembling and incorporating the information from surroundings. It helps them to predict the forward behavior of the network based on the current states. The performance parameters observed in a network node are collected and uploaded into the network by each cognitive element for decision-making. The decision-making process uses reasoning to determine the next set of actions that can be implemented in the network. Learning and predicting operation in a router depends upon the QoS parameters considered. For this purpose, all the router's forward conditions are sent to its neighboring routers. Information is shared among routers by sending Cognitive Packets as in Gelenbe (2001). These cognitive packets demand negligible channel bandwidth compared to the regular packets carrying information, thus minimizing the demand of capacity for cognition activity in the network. At this point when the user wishes to forward the data to a particular end user, the information about other nodes connected from the sender to the rest of the participating nodes will be collected by the intelligent nodes. These sets of nodes participate in the selection process of graphical node representation. If the nodes are selected then GA is applied for path selection. From this the best path will be selected based on the shortest hop count.

The graded network as defined in Nair (2011) is a network with environment awareness. The efficacy and availability of the nodes as determined by the grading approach produces the feasibility of the nodes to forward the data. Here in this paper the QoS parameters considered are delay, jitter and loss thus creating the environment awareness and thus obtaining the optimal path based on available bandwidth.

## 3 QoS and Reasoning

Controlling congestion so as to guarantee QoS for the flow in a network is a challenging task. Congestion could have occurred due to many reasons especially due to the workload being high at the routers and the demand being high for the channels. *This may be due to the fact that buffer is full which leads to QoS deteriorating and loss of packet cause of buffer being full*. It also increases the average packet queuing time. In wireless network, it is further complicated, due to the presence of noise and nodes mobility. As a consequence, wired and wireless are integrated together, the analyses and prediction of the behavior of the network becomes difficult. This complexity has risen due to the fact that the network is not aware of its sate and its needs, will not have knowledge of its goals and how to achieve them, and finally will not be able to reason for its actions. The drive for Cognitive Network is reasoning and cognition capability. This will enable the construction of dependable, resilient, self-manageable, self-configurable and finally survivable networks. The diagnosis decisions will lead to many solutions. Now the networks require a capability for acting and healing to sustain at the best operations. The realization of acting and healing also depends on:

- Building dynamic active resource which can support network whenever required to install/ uninstall a new service.
- Managing and coordinating the active service modules
- Defining a path that is learnt information, without knowing the details of physical network.

However in general, solutions need to meet network performance quality and service level agreement profiles as well as cost profiles. A better quality in network servicing through autonomic means is helpful to limit the administrator intervening in the dynamic, tedious, error prone and trivial tasks of network management. It would be better to give more strategic tasks like setting high-level goals and policies, monitoring and training of the cognition layer etc. to the administrator.

For cognition to work effectively, one has to consider administrative policies, resource availability, user needs, and service level agreements, too. In simple analytical solution in case conflict arises one can solve by the list of prioritize to define the goal. But this will not be an efficient way for conflict solving in case there are varying number of users and a set of varying coefficient (i.e. parameter metrics). An appropriate representation for the parameter would be,

- Active services
- Policies
- Resources
- Network characteristic
- User request.

Active services are communications taking place on demand satisfying the requests of the user. Policies are the rules to be followed in the Internet or future public net. Resources are available bandwidth and buffer management specified in the intermediate nodes. Network characteristics are emphasized with QoS and considered for the design. User request is the demand from the user.

This offers the reasoning unit with necessary input, which also can be the input to the intelligent system that may implement a role-based agenda to facilitate both decision-making and conflict resolution. As a last responsibility, cognitive system needs to be able to diagnose problems, classify them, and take appropriate actions to recover from them. In case of insufficient knowledge it must be able to take or query the administrator.

The cognitive process must monitor (Observe) sustainability based on a set of resources. The results are evaluated based on the flowchart (Plan). Diagnostic decisions are taken in accordance to the predefined goals based on expectation and QoS. These two factors can be given by weighted functions defined in a Hidden Markov Model Nair (2008). The diagnostic and operational decisions will trigger healing solutions workflows (Act), which will result in all possible solutions for qualitative and cost effective network performance. The healing process is more complex when more than one solution is available for any one network problem. The solution's selection must be very concise and precise so as to have the solution space smaller. We can achieve the smaller space through reasoning and set theory.

Among the reasoning techniques available in AI, abductive reasoning contributes to cognitive based assessment in three different approaches. Firstly, it explains the alternate paths for understanding the conceptions and misconceptions. Secondly, it is developed based on reverse engineering, there by developing a model from the end goal. Third, analogical reasoning is inseparable from it, as the idea of learning is from pattern recognition and suggested hypotheses. Thus abduction is a type of critical thinking, than just symbolic logics, which has found its application in cognitive networks.

Considering the representation in mathematical form, we see that classical set theory has found its application in many fields. Because of the assumptions made in it, we see its application is limited in few areas. Instead we can use the rough sets theory which incorporates the model of knowledge into its formalism, thus it represents sets approximately in terms of the available context knowledge, and it leads to approximate decision due to the imperfect context knowledge. The rough set theory classifies the contents into imprecise, uncertain or incomplete information. It is a very effective methodology for data analysis and discovering rules in the attribute-value based domains. Here we define a set, which contains domain objects of particular attributes, so that the goal of concept learning is to find the discriminating description of object with a specific decision attribute set. This is done so as to be specific in reaching the goal. Planning how to deal with multiple goals, using same resources is also taken care by this approach.

## 4 Genetic Algorithms

Genetic algorithms are a part of evolutionary computing. It is also an efficient search method that has been used for path selection in networks. These stochastic search algorithms are based on the principle of natural selection and recombination. GA has been an efficient search method based on principles of natural selection and genetics. They are being applied successfully to find acceptable solutions to problems in business, engineering, and science. We can find good solution for adequate amount of data at hand, but the complexity of data increases as GA takes time to find the solution. GA works well for network model to find the optimal path. In this, the source and the destination nodes are sure to participate in every generation. Other nodes or the genes become a part of the chromosome if they find an optimal path between the source and destination.

GA is composed with a set of solutions, which represents the chromosomes. This composed set is referred to population. Population consists of set of chromosome which is assumed to give solutions. From this population, we randomly choose the first generation from which solutions are obtained. These solutions become a part of the next generation. Within the population, the chromosomes are tested to see whether they give a valid solution. This testing operation is nothing but the fitness functions which are applied on the chromosome. Operations like selection, crossover and mutation are applied on the selected chromosome to obtain the progeny. Again fitness function is applied to these progeny to test for its fitness. Most fit progeny chromosome will be the participants in the next generation. The best sets of solution are obtained using heuristic search techniques. The general description of GA is as follows:

a. **First Generation** randomly pick n chromosome to form a population assuming that this could be the probable solution to the problem.

b. **Fitness Function** the fitness function f(x) is applied on each chromosome in the generation.

c. **Next Generation** create the next generation by performing the following steps until *n* chromosomes are obtained

   i. **Selection operation** Select any two best fit chromosome from the generation
   ii. **Crossover** with a defined probability apply the crossover technique for the above obtained chromosome to form the children
   iii. **Mutation** with a defined probability mutate a new gene at desired position
   iv. **Test** whether the obtained children are fit to go to next generation. If yes, then move them to next generation.

d. **Test** if generation is of desired size, if yes, stop, and return the best solution from the current generation.
e. **Repeat** go back to b

The performance of GA is based on efficient representation, evaluation of fitness function and other parameters like size of population, rate of crossover, mutation and the strength of selection. Genetic algorithms are able to find out optimal or near optimal solution depending on the selection function as described by Goldberg (1995) and Pal (2004).

*4.1 Proposed Algorithm*

The proposed algorithm follows the above GA steps to obtain the optimal path. The input to the path selection GA is the set of nodes which would satisfy three QoS parameters (delay, jitter and loss). This kind of selection is made possible by sending information in the packet of the node itself as given by Nair (2008). In the next stage depending on bandwidth availability, path is obtained using GA. From this set of nodes, the initial population of nodes is chosen at random which forms the chromosome. For these selected nodes we would calculate the available bandwidth ($A_b$) using the formula,

$A_{b_{j=1}^m}$ = link utility- required bandwidth by the data to be sent.

$$A_{b_j^m = 1} = Link_{utility} - Required\ Bythenode \quad (1)$$

The link utility is stored in a vector and is referred for calculation in $A_b$. If $A_b > 0$, then the link can participate in the optimal path, otherwise it is not chosen. A reference chromosome starting at source and ending at destination will always be selected in every population by elitism. The fitness of the chromosome is calculated using,

Path selection $f_j(t)$ = Available bandwidth/ summation of $i^{th}$ chromosome.

$$f_j(t) = \frac{Ab_j}{\sum_{i=1}^m A_{b_i}(t)} \quad (2)$$

The probability of selecting the chromosome $f_j(t)$, is given by,

$$P_j(t) = \frac{f_j(t)}{\sum_{i=1}^m f_j(t)} \quad (3)$$

If $f_j(t)$ is between 0.5 and 1 those chromosomes gets selected and data is forwarded in that path after the convergence of the generations. The algorithm is as follows:

**begin** PATHSELECTION_GA

    Create initial population of n nodes randomly.

    **while** generation_count < k do

    /* k = max. no. of generations.*/

    **begin**

    Selection

    Fitness Function

    Modified crossover

    Mutation

    Increment generation_count.

  **end** ;

Output the optimal path by selecting the highest probability value chromosome on which data can be sent

**end** PATHSELECTION_GA.

Initially, we have applied the QoS parameters like delay, jitter and loss which are considered affecting the node's participation in routing as in Nair (2010). Node delay is calculated based on the following formula with fixed message size ($M_{Size}$) and random Bandwidth ($B_i$) value,

$$\text{Delay} = \frac{M_{size}}{B_i} \tag{4}$$

Here the propagation delay, transmission delay and the processing delay were neglected as the simulation setup designed would not encounter any such additional delay and reality is reflected in the QoS parameter set used. Jitter and loss are random values assigned. The selection of the nodes with respect to these three parameters was considered when the node's parametric value was below a threshold value. The selection of threshold value is a predetermined pattern for the simulation. This will help the convergence to destination node fast.

### *4.2 Architecture Model*

The proposed work is based on the model as shown in Fig.1. This model is based on the concept proposed by Mitola, which is given by Thomas (2005), where the input to the cognitive controller is the data or information collected about the nodes from the network by intelligent agents. These agents can be intelligent routers, sensors or any form of cognitive packet makers where the status of the nodes is collected Nair (2011). The controller then forwards the information to the map developed module, where the graphical representations about the complete nodes participating are formed. The selections of these nodes are based on the QoS parameters considered. On this map the genetic algorithm technique is applied to obtain the optimal paths. From the generations we derive the optimal path based on the shortest paths. This information is stored in the knowledge base for further reference, which can be used later if such a scenario reoccurs.

### *4.3 Representation*

The network under consideration is represented as G = (V, E), a connected non-loop free graph with N nodes. The metric for optimization is the bandwidth available between the nodes. The goal is to find the path with availability of bandwidth between source node $V_s$ and destination $V_d$, where $V_s$ and $V_d$ belong to V. E is the set of edges connecting the nodes which are represented in V. From this topology we develop a graphical representation of QoS satisfied by the nodes and generation of optimal path using genetic algorithm. Finally data is sent along the generated path.

### *4.4 Natural Selection*

GA uses a selection mechanism to select the individuals from the population to insert into the mating pool. Individuals from the mating pool are used to generate new off spring which will participate in the next generation. As the individuals of the next generation are going to participate further this is better for the genes to be of good condition. The selection mechanism is a process of selecting the individuals of good condition. This selection function leads to a better population with good condition. The convergence rate largely depends on the selection function.

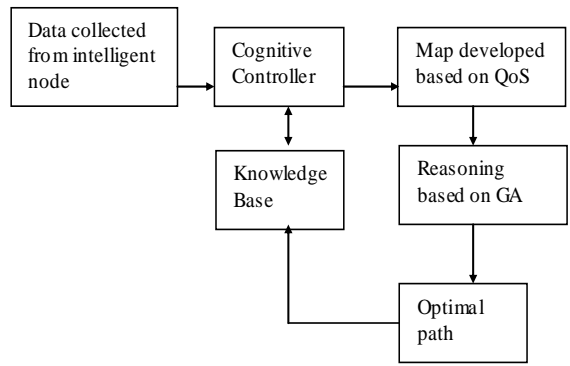

Figure 1. Cognition Model

Here the selection process is carried out by Roulette Wheel method. In this method, the individuals are chosen based on the relative fitness with its competitors. Here the wheel is divided into slice, where the fit chromosome gets larger slice. For selecting the chromosome for next generation the wheel is spun. Once the wheel stops, the individual corresponding to the slice on which it lands goes to next generation.

*4.5      Singlepoint / Multipoint Crossover*

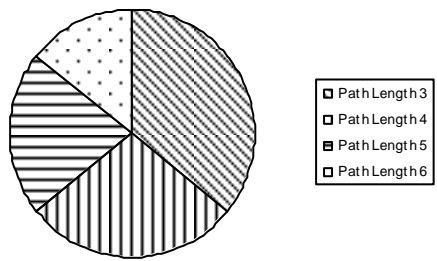

Figure 2. Roulette Wheel

Crossover operator combines sub parts of two parent chromosomes and produces off spring that contains some parts of both the parent. In single point crossover technique, one offspring consists of first part of one parent and second part of the other parent. Similarly the other offspring is generated. Here we also use multi point crossover mechanism called partially mapped crossover. In this two chromosomes are picked at random. The strings between the crossover sites are exchanged position by position; other elements are determined by ordering information, which is partially determined by each of its parents.
For example:

Single point crossover:

Parent 1 :1 2 3  4 | 5  6 7

Parent 2 :1 3 4 6  | 2 5 7

Offspring 1 :1 3 5 6 2 5 7

Offspring 2 :1 2 6 4 5  6 7

Multipoint crossover:

Parent 1 :1 2 3 | 4 5 | 6 7

Parent 2 :1 3 4 | 6 2 | 5 7

Offspring 1 :1 3 5 6 2 4 7

Offspring 2 :1 2 6 4 5 3 7

Here in this paper a multi point crossover technique has been applied for all equal length chromosomes. We have chosen breadth first search technique to determine all possible paths from source to destination. From this output another file was generated which grouped the paths of same length in ascending sequence. Later the chromosome selections were made from this pool of population where same length paths were considered for generating the children using crossover technique. The example given here indicated nodes who participated in routing which formed a path from source to destination. For example, in Parent1, '1' is the source which is connected to node number '2' as so on until destination node '7' is reached.

### 4.6  Mutation

Sometimes it may be possible that by crossover operation, a new population never gets generated. To overcome this limitation, we do mutation operation. Here we use insertion method, as a node along the optimal path may be eliminated through crossover.

## 5  Simulation and Results

The initial configuration of simulation model consists of a randomly distributed set of nodes, in a geographical area. They are capable of getting connected through links which could be created on demand. The information on the topology link connectivity is made available in a file. Based on this information initially BFS is run to derive all possible paths from source to destination and is put in a sort sequence based on the length of the chromosome and made available in another file. Later GA is applied on this based on the four QoS metric where the delay, jitter and loss were compared and the bandwidth of the link were determined by the objective function. The nodes with low delay, jitter and loss were selected.  This would reduce the population size we were looking at if nodes do not satisfy the threshold value for the three QOS metric. In this reduced population, GA was applied. The chromosome or the path is selected by elitism process, which would be the shortest path obtained by the BFS algorithm. Remaining of the children are derived by the crossover technique which is applied 95% of the simulation. 5% of the simulation time is assigned to the mutation technique. The selection of the chromosome is based on the bandwidth property being satisfied. If the probability value of the bandwidth being satisfied is above threshold value then the path gets to participate in next generation. This process is repeated until the fixed number of generation is reached. The values obtained in the last generation, determines which among the chromosome will get selected as the optimal path. In the final generation the chromosomes / paths which have the probability value above 0.5 and hop count been less will be selected for optimality.

In a cluster of network nodes where GA method was applied for optimizing the path, a propagation of generation and its details are given below. The topology of one random run with ten nodes in the network is shown in Fig. 3. Here the paper distinguishes itself by deciding the channels based on bandwidth availability using GA. The result obtained is based on the best bandwidth available at minimum hop count. Initially ten random chromosomes are generated and placed in the roulette wheel based on the path length. Shorter the path length higher is the probability of selecting the chromosome from the roulette wheel. Out of the ten random chromosomes generated five best are considered for the first generation. Here the result is shown for five such generations which are obtained by applying crossover and insertion mutation functions. At each generation of population, validation of the chromosome is carried out and the best fit chromosome is only considered for next generation. It was found that probability of shortest path convergence was faster. The results obtained are shown in the table for the five generations.

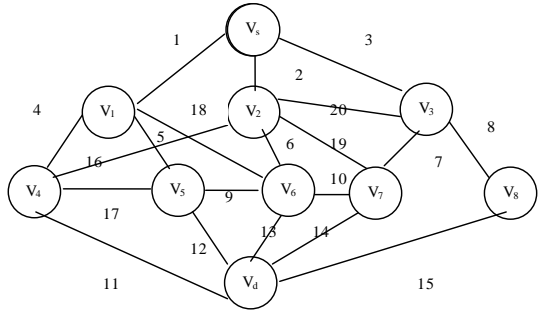

Figure 3: QoS Selection Map

The first generation input was chosen randomly based on roulette wheel selection method, for which fitness function is applied and probability of its survival has been calculated as shown below:

TABLE I. GENERATION 1

| Chromosome | No. of Nodes Visited | Fitness | Probability of selecting chromosome |
|---|---|---|---|
| C1 | 3 | 0.2903 | 1 |
| C4 | 4 | 0 | 0 |
| C3 | 3 | 0.3030 | 0.5106 |
| C8 | 5 | 0.1766 | 0.2294 |
| C7 | 5 | 0 | 0 |

From the first generation we see that crossover probability on any two chromosome of equal length does not produce any children. So insertion mutation was applied on C3 chromosome to obtain C6 chromosome. By elitism C1 chromosome exist and remaining are obtained by the selection method. The results are as in table 2.

TABLE II. GENERATION 2

| Chromosome | No. of Nodes Visited | Fitness | Probability of selecting chromosome |
|---|---|---|---|
| C1 | 3 | 0.2903 | 1 |
| C5 | 4 | 0.2045 | 0.4132 |
| C6 | 4 | 0.2173 | 0.3052 |
| C9 | 6 | 0.1267 | 0.1510 |
| C10 | 6 | 0 | 0 |

The third generation children are obtained from the multipoint crossover technique on C9 and C10 chromosomes to obtain C11. The remaining chromosomes are obtained as in second generation. Similarly the fourth generation was obtained by crossover on C1 and C3 to obtain C12. The fifth generation chromosome did not produce any valid paths after crossover technique and since it was the last generation and mutation operation was to be applied only 0.01 percent of time, the chromosome was selected based on selection method. The results are shown in the tables below.

TABLE III. GENERATION 3

| Chromosome | No. of Nodes Visited | Fitness | Probability of selecting chromosome |
|---|---|---|---|
| C1 | 3 | 0.2903 | 1 |
| C11 | 6 | 0 | 0 |
| C9 | 6 | 0.1267 | 0.3038 |
| C3 | 3 | 0.3030 | 0.4208 |
| C7 | 5 | 0 | 0 |

TABLE IV. GENERATION 4

| Chromosome | No. of Nodes Visited | Fitness | Probability of selecting chromosome |
|---|---|---|---|
| C1 | 3 | 0.2903 | 1 |
| C12 | 3 | 0.3214 | 0.5253 |
| C7 | 5 | 0 | 0 |
| C8 | 5 | 0.1766 | 0.2240 |
| C5 | 4 | 0.2045 | 0.2059 |

TABLE V. GENERATION 5

| Chromosome | No. of Nodes Visited | Fitness | Probability of selecting chromosome |
|---|---|---|---|
| C1 | 3 | 0.2903 | 1 |
| C7 | 5 | 0 | 0 |
| C8 | 5 | 0.1766 | 0.3782 |
| C10 | 6 | 0 | 0 |
| C2 | 3 | 0 | 0 |

From the result obtain, we apply the minimum path algorithm to obtain the optimal path. The criteria of this algorithm are to see that it has to traverse through less number of nodes and must possess higher probability value greater than 0.5. Thus we find that path length of three which is the shortest path, having higher probability, has been selected for optimal path selection.

## 6 Conclusions and Future Work

This work presented an optimal path selection technique using genetic algorithm. The network routing performance was enhanced using the intelligent routing approaches. Here the data was forwarded based on the bandwidth availability and path selection achieved through the dynamic GA model. The results show a better convergence of path selection for shortest length chromosome.

Here the best selection of the path was obtained only based on shortest hop count. This can be improved further based on geographical coupling, where we can collect and connect the geographic information, translating the information of destination into geographical measures also for judging the route selection at one stage or several stages.

## References

T. R. Gopalakrishnan Nair, Kavitha Sooda (2011)" Application of Genetic Algorithm on Quality Graded Networks for Intelligent Routing", IEEE World Congress on Information and Communication Technologies, pp. 558-563, DOI: 10.1109/WICT.2011.6141306.

Erol Gelenbe (2006), "Genetic algorithms for route discovery", *IEEE Transaction on Systems, Man and cybernetics*, vol. 36, No.6, pp. 1247-1254.

Cauvery N K, K V Viswanatha (2009), " Routing in dynamic network using ants and genetic algorithm", *IJCSNS*, vol.9, No.3, pp. 194-200.

Elizabeth M Royer, Chai-keong Toh (1999), "A review of current routing protocols for adhoc mobile wireless networks", *IEEE Personal Communication*, pp. 46-55.

AliSelamat, Sigegu Omatu (2003),"Analysis on route selection by mobile agents using genetic algorithm", *SICE Conference* in Fuki, Japan, Vol. 2, pp. 2088-2093.

Aluizio FR Araujo, Maury m Gouvea Jr (2006), "Multicast routing using genetic algorithm seen as a permutation problem", *AINA Proc., IEEE Computer Society*, Vol. 1, pp 6.

Marios P Saltouros, Maria E Markaki, Anastasios K Taskaris, Michael E Theologou, Iakovos S Venieris (2000), " A new route selection approach using scaling Techniques: An application to hierarchical QoS-Based routing", *LCN 2000 Proc.*, pp. 698-699.

Hitoshi Kanoh, Tomohiro Nakamura (2000), "Knowledge Based Genetic Algorithm for Dynamic Route Selection", *International Conference a knowledge-Based Intelligent Engineering Systems & Allied Technologie*, pp. 616-619.

M.Roberts Masillamani, Avinankumar Vellore Suriyakumar, Rajesh Ponnurangam, G.V.Uma (2000), " Genetic algorithm for distance vector routing technique", *AIML 06 International Conference*, pp. 160-163.

Luiz A DaSilva, Allen B. Mackenzie, Claudio R C, M DaSilva, Ryan W Thomas (2008), "Requirements of an Open Platform for Cognitive Networks Experiments", *DySPAN Symposium, IEEE*, pp. 1-8.

Xiliang Liu, Kaliappa Ravindran, Pmitri Loguinov (2008), "A Stochastic Foundation of Available Bandwidth Estimation: Multi-Hop Analysis", *IEEE Transactions on Networking*, VOl.16, No.1, pp. 130-143.

C P Lokuge, A J Coulsom, J Gao (2009), "A Novel Geographic Set-up and an Access Protocol for Mesh, Ad-Hoc and Cognitive Networks", *ICCMC*, pp. 234-240.

Nolan, Rondeau T W, Sutton, Bostain, Doyle (2007), "A Framework For Implementing Cognitive Functionality", *IEEE Proc*, pp. 1149-1153.

Neeli Rashmi Prasad (2008), "Secure Cognitive Networks", *EuWit European Confeece* pp. 107-110.

Paul Sutton, Linda E Doyle, K E Noelan (2006), "A Reconfigurable Platform for Cognitive Networks", *International Conference on Cognitive Radio Oriented Wireless Networks and Communications,* pp. 1-5.

T.R.Gopalakrishnan Nair, Abhijith, Kavitha Sooda (2008), "Transformation of Networks through Cognitive Approaches", *JRI- Journal of Research & Industry* Volume 1 Issue 1, pp.7-14, December.

Erol Gelenbe (2001), "Cognitive Packet Networks," *Proceedings of the 11th IEEE International Conference on Tools with Artificial Intelligence*, vol 46, pp.155-176.

Brad l Miller, David E Goldberg (1995), " Genetic algorithm, Tournamanet selection and effect of noise", *Journal Computer Systems*, vol. 9, pp. 193-212.

Shubhra Sankar Ray, Sanghamitra Bandyopadhyay and Sankar K. Pal (2004), "New Operators of Genetic Algorithms for Traveling Salesman Problem", *ICPR 2004*, pp. 129-170.

T R GopalaKrishnan Nair, M Jayalalitha, Abhijith S (2008), "Cognitive Routing with Stretched Network Awareness through Hidden Markov Model Learning at Router Level", *IEEE Workshop on Machine learning in Cognitive networks*, Hong Kong.

Ryan W Thomas, Luiz A DaSilva, Allen B Mackenzie (2005), "Cognitive Networks", *Proceedings of IEEE DySPAN 2005*, pp. 352-360.

T. R. Gopalakrishnan Nair, Kavitha Sooda (2010), "A Novel Adaptive Routing through Fitness Function Estimation Technique with multiple QoS parameters Compliance", ICIP-2010 Conference in Bangalore, August 2010.